\documentclass[traditabstract]{aa}
\usepackage{graphicx}
\usepackage{txfonts}
\usepackage{natbib}
\usepackage{rotating}           
\bibpunct{(}{)}{;}{a}{}{,}

\newcommand{\tempo}{\textsc{tempo2}}
\newcommand{\msun}{M$_\odot$\,}

\newcommand{\kms}{km\,s$^{-1}$}

\begin{document}

  \title{Long-term timing of four millisecond pulsars}

  \author{G.\ H.\ Janssen\inst{1,2}
    \and B.\ W.\ Stappers\inst{1,2,3}
    \and C.\ G.\ Bassa\inst{1,4,5}
    \and I.\ Cognard\inst{6}
    \and M.\ Kramer\inst{7,1}
    \and G.\ Theureau\inst{6,8}}

  \institute{Jodrell Bank Centre for Astrophysics, School of Physics and Astronomy,
 University of Manchester, Manchester M13 9PL, UK\\ {email:gemma.janssen@manchester.ac.uk}
  \and Astronomical Institute ``Anton Pannekoek'', University of
    Amsterdam, Postbus 94249, 1090 GE Amsterdam, The Netherlands;
  \and Stichting ASTRON, Postbus 2, 7990 AA Dwingeloo, The Netherlands
  \and SRON, Sorbonnelaan 2, 3584 CA Utrecht, The Netherlands
  \and Department of Astrophysics/IMAPP, Radboud University Nijmegen,
  PO Box 9010, 6500 GL Nijmegen, The Netherlands
  \and Laboratoire de Physique et Chimie de l'Environnement, CNRS, 3A
  Avenue de la Recherche Scientifique, F-45071 Orl\'eans, Cedex\,2,
  France
  \and Max-Planck-Institut f\"ur Radioastronomie, Auf dem H\"ugel 69, 53121 Bonn, Germany
  \and Galaxies, \'Etoiles, Physique, Instrumentation, CNRS URA 1757, 92195 Meudon Principal Cedex, France}

\date{Received/Accepted}

     \abstract{

       We have timed four millisecond pulses, PSRs\,J1721$-$2457,
       J1745$-$0952, J1810$-$2005, and J1918$-$0642, for up to a total
       of 10.5 years each using multiple telescopes in the European
       Pulsar Timing Array network: the Westerbork Synthesis Radio
       Telescope in The Netherlands, the Nan\c cay Radio Telescope in
       France and the Lovell telescope at Jodrell Bank in the UK.  The
       long time span has enabled us to measure the proper motions of
       J1745$-$0952 and J1918$-$0642, indicating that they have
       transverse velocities of 200(50) and 54(7)\,\kms
       respectively. We have obtained upper limits on the proper
       motion of J1721$-$2457 and J1810$-$2005, which imply that they
       have transverse velocities less than 140 and 400\,\kms
       respectively. In all cases, the velocities lie in the range
       typical of millisecond pulsars.  We present pulse profiles for
       each pulsar taken from observations at multiple frequencies in
       the range of 350 to 2600 MHz, and show that J1810$-$2005 shows
       significant profile evolution in this range. Using our
       multi-frequency observations, we measured the spectral indices
       for all four pulsars, and for J1810$-$2005 it appears to be
       very flat.  The flux density of J1918$-$0642 shows extensive
       modulation which we attribute to the combined effects of
       refractive and diffractive scintillation.  We discuss the
       possible use of including J1721$-$2457 or J1918$-$0642 in a
       pulsar timing array, and find that J1918$-$0642 will be useful
       to include when the timing precision of this pulsar is improved
       over the next few years. We have searched archival optical
       observations to detect companions of the binary pulsars, but
       none were detected. However, we provide lower limits on the
       masses of the white dwarf companions of PSRs J1745$-$0952 and
       J1918$-$0642.

}

\maketitle

\section{Introduction}

In this paper we present improved timing solutions for four
millisecond pulsars (MSPs): PSRs J1721$-$2457, J1745$-$0952, and
J1918$-$0642, discovered by \cite{eb01b}, and J1810$-$2005,  
that was discovered by \cite{clm+01}.
Long-term timing of MSPs is an important tool to determine masses of
the individual stars in binary systems, and by constraining secular
effects like proper motion, it can be used to improve statistics on
the evolution of these systems.

In general, masses of pulsars are not easy to determine.  In some
cases, when the pulsar is in a binary with another compact object,
high precision pulsar timing observations on extended timescales can
allow for detecting post-Keplerian parameters of the systems, which
can be used to separately measure the individual masses of the stars.
    Space velocities derived from proper motion measurements of radio
    pulsars give clues about the evolution of these systems and their
    birth supernovae (e.g.\,\cite{tsb+99,hllk05,lor08}).  It is
    believed that recycled pulsars have lower space velocities than
    normal, slowly rotating pulsars.  However as MSPs are generally
    the most stable rotators, the effects of their space velocities
    on their rotational and orbital parameters will be easier to derive
    from timing measurements of those systems and could be used to
    deduce the intrinsic properties of the stars. 

Another interesting and important use of timing 
MSPs to high precision, is the formation of a pulsar timing array
(PTA, e.g. \citealt{hjl+08a,jsk+08a}). An instrument like this
  will use an array of MSPs as the endpoints of a Galaxy-scale
  gravitational wave (GW) detector.  Current estimates predict that to
  detect a GW background, long-term high precision timing of about
  20\,MSPs is required \citep{jhlm05,vlml08}.  Increasing the number
  of stable MSPs in the array will improve the detection significance.
  In order to better understand these systems in general, and to
  determine their suitability for inclusion in a timing array, their
  long-term timing behaviour needs to be determined. Moreover, the
  best frequency at which they should be timed needs to be
  ascertained. This is a combination of the pulse shape at these
  frequencies and their intensity. We observed all four pulsars at
  additional frequecies to find their best possible observing
  frequency for timing purposes, and discuss their suitability for
  inclusion in a PTA.

Three of the pulsars that are presented in this paper are in
low-eccentricity binary systems with white dwarf companions (see
Table\,\ref{tab:solution}).  These systems are usually classified as
low-mass binary pulsars (LMBPs).  The LMBPs distinguish themselves
from intermediate-mass binary pulsars (IMBPs) in having shorter
periods ($<10$~ms), very low eccentricities ($<10^{-5}$), and they
follow relationships between their orbital period and the eccentricity
of their orbit, and their orbital period and the companion mass
\citep{tc99}.  The origin of the difference between these two classes
of binary MSPs is believed to be an evolutionary effect, and mainly
due to the progenitor masses of the companion stars.  There are now
$\sim$65 MSP binary pulsars with probable white dwarf companions known
(e.g.\,\citealt{lor08}) and about 16 of these can be regarded as IMBP
candidates \citep{jcv+06}.

Detecting an optical counterpart to one of the binary pulsars allows
to derive properties of their white dwarf companions.
Because of differences in the cooling properties of white dwarfs in
LMBPs and IMBPs, optical observations 
can in some cases be used to distinguish between the two classes
(e.g.\,\citealt{vk95}).
 So far, for all three binary pulsars described in this
paper, the most recently published limit on optical magnitude of the
companion is $R>24$ \citep{vbjj05}.

\section{Pulsar timing observations and data analysis}\label{s:timing}

\subsection{Westerbork}\label{s:wsrt}

The pulsars were observed approximately monthly using the pulsar
machines at the Westerbork Synthesis Radio Telescope (WSRT): PuMa
\citep{vkh+02} since August 1999, and PuMaII \citep{ksv08} since
August 2006.  The WSRT has three frequency ranges allocated for pulsar
timing; for observations centred at 350\,MHz we use two bands of
10~MHz wide, and observations centred at 1380 or 2300\,MHz use 80\,MHz
of bandwidth, spread in 8 steps of 10\,MHz over a range of
160\,MHz. The new pulsar machine uses the full 80\,MHz bandwidth at
the low frequencies, and all of the available 160\,MHz bandwidth at
1380 and 2300\,MHz.
For all four pulsars presented here, we mainly used the 1380\,MHz band
as the pulsars were detected in this band, and there are no other
reports of detections in other bands. Also this is the best available
band for timing at the WSRT as it has the best sensitivity. For
completeness, we observed all pulsars for one hour at 350 and
2300\,MHz, see Fig\,\ref{fig:profiles}.

For a complete description of the PuMa data analysis, we refer to
\cite{jsk+08}.  Analysis of the PuMaII data was carried out using the
PSRCHIVE \citep{hvm04} software suite.  For both instruments, each
profile was cross-correlated with a standard profile at the
corresponding observing frequency (Fig.\,\ref{fig:profiles}), obtained
from the summation of high signal-to-noise (S/N) profiles, to
calculate a time of arrival (TOA) for each observation. These were
referred to local time using time stamps from a H-maser at WSRT and
converted to UTC using global positioning system (GPS) maser offset
values measured at the observatory, and GPS to UTC corrections from
the Bureau International des Poids et Mesures
(BIPM)\footnote{http://www.bipm.org}. The TOAs were
converted to the Solar system barycentre using the JPL ephemeris
DE405\footnote{ftp://ssd.jpl.nasa.gov/pub/eph/export/DE405/de405iom.ps}.
We used the \tempo\ timing software package \citep{hem06} to analyse
the data.

Flux densities were calculated for the WSRT observations by using the
aforementioned bandwidth and observing times, the measured
S/N ratio for each profile, the gain of the WSRT
(1.2\,K/Jy), the pulse duty cycle ($\sim$10\%), the system temperature
and the pulsar specfic radiometer equation \citep{dtws85}. The total
system temperature was determined from the synthesis data that is
taken in parallel with the pulsar observations.

\subsection{Nan\c cay}

PSR\,J1721$-$2457, J1745$-$0952 and J1918$-$0642 were observed roughly
every 3 to 4 weeks with the Nan\c cay Radio Telescope (NRT) since 2006.

Equivalent to a 93-m dish, the Nan\c cay Radio Telescope and the BON
(Berkeley-Orl\'eans-Nan\c cay) coherent dedispersor were used for
typical integration times of 45 min.  Coherent dedispersion of a 64
MHz band centred on 1398 MHz was carried out on sixteen 4 MHz channels
using a PC-cluster in the period covered by the observations.
  The Nan\c cay data are written with timestamps which are directly
  tied to UTC(GPS) in realtime, using a Thunderbolt receiver (Trimble
  Inc.), when converting from an analogue to digital signal is
  performed.
Differences between UTC and UTC(GPS) are less than 10~ns at
maximum, and therefore no laboratory clock corrections are needed. One
TOA was calculated from a cross-correlation with a pulse template for
each observation of $\sim$~45 min.

The flux densities for the Nan\c cay data were calculated in a similar way to
those for the WSRT except that the system temperature was determined
from the known receiver temperature and sky temperature in the
direction of the source. The gain of the NRT is 1.4\,K/Jy.

\subsection{Jodrell Bank}

PSR\,J1810$-$2005 was observed at Jodrell Bank since November
1997, and PSR\,J1918$-$0642 since February 2008. Both pulsars were
observed every two weeks at a centre frequency around 1400\,MHz.
The gain of the Lovell telescope is 1\,K/Jy.

The polarized signals from the receiver were were fed into an analogue
filterbank system with $2\times 32\times 1$\,MHz channels, and
filtered and digitised at appropriate sampling intervals and
incoherently dedispersed in hardware.  The resulting dedispersed
timeseries were folded on-line with the topocentric pulsar period and
finally written to disc. In the off-line reduction, the two
polarisations were summed to form total-intensity profiles. A standard
high S/N pulse template was used to determine a TOA for each
observation. During this process, TOAs were referred to the local
H-maser time-standard and already corrected to UTC using information
obtained via the GPS.

\section{Pulsar timing results and discussion}

\begin{table*}
  \caption{Timing solutions for the pulsars.
    \label{tab:solution}}
\begin{tabular}{lcccc}
\hline\hline \\*[-2ex]
Pulsar name\dotfill &  J1721$-$2457 & J1745$-$0952 & J1810$-$2005 & J1918$-$0642\\ 
\hline \\*[-2ex]
\multicolumn{2}{c}{Fit and data-set} \\
\hline \\*[-2ex]
MJD range\dotfill &  52076--54799 & 52076---54776 & 50757--54796 & 52094-54814\\ 
Number of TOAs\dotfill & 120 & 113 & 491 & 152 \\
Rms timing residual ($\mu$s)\dotfill & 31.9 & 20.8 & 270.7 & 2.6\\
Weighted fit\dotfill &  Y & Y & Y & Y \\ 
Reduced $\chi^2$ value$^{a}$\dotfill & 1.05 & 1.02 & 1.01 & 1.04 \\
\hline \\*[-2ex]
\multicolumn{2}{c}{Measured Quantities} \\ 
\hline \\*[-2ex]
Right ascension, $\alpha$ (J2000)\dotfill & 17$^\mathrm{h}$21$^\mathrm{m}$05\fs4963(3) & 17$^\mathrm{h}$45$^\mathrm{m}$09\fs1348(2) & 18$^\mathrm{h}$10$^\mathrm{m}$58\fs9919(6) & 19$^\mathrm{h}$18$^\mathrm{m}$48\fs035270(13)\\ 
Declination, $\delta$ (J2000)\dotfill & $-$24\degr57\arcmin06\farcs36(6) & $-$09\degr52\arcmin39\farcs682(15) &  $-$20\degr05\arcmin08\farcs27(14)& $-$06\degr42\arcmin34\farcs8636(6)\\ 
Pulse frequency, $\nu$ (s$^{-1}$)\dotfill & 285.989343507338(15) & 51.609431233273(5) & 30.467142155106(7)& 130.7895141841235(6) \\ 
First derivative of pulse frequency, $\dot{\nu}$ (s$^{-2}$)\dotfill & $-$4.533(5)$\times 10^{-16}$ & $-$2.4627(13)$\times 10^{-16}$ & $-$1.3684(11)$\times 10^{-16}$& $-$4.39525(17)$\times 10^{-16}$ \\ 
Dispersion measure, DM (cm$^{-3}$pc)\dotfill & 48.34(3) & 64.27(9) & 241.0(3) & 26.554(10) \\
Proper motion in RA, $\mu_\mathrm{\alpha}$ (mas\,yr$^{-1}$)\dotfill & 1.8(1.8) & $-$21.2(1.1) & 0(2) & $-$7.20(10)\\
Proper motion in DEC, $\mu_\mathrm{\delta}$ (mas\,yr$^{-1}$)\dotfill & $-$14(25) & 11(5) & 17(37)& $-$5.7(3) \\
Binary\dotfill & $-$ & ELL1 & ELL1 & ELL1 \\
Orbital period, $P_\mathrm{b}$ (d)\dotfill & & 4.943453386(12) & 15.01201911(4) & 10.9131777486(12) \\
Projected semi-major axis of orbit, $x$ (lt-s)\dotfill & & 2.378599(5) & 11.977880(18) & 8.3504716(4) \\
TASC (MJD)\dotfill & & 53198.621445(3) & 53195.528458(4) &  53402.53135510(8)\\
EPS1 ($\epsilon_1$)\dotfill & & \phantom{$-$}0.000009(3) & \phantom{$-$}0.000009(3) & $-$0.00001244(9) \\
EPS2 ($\epsilon_2$)\dotfill & & $-$0.000004(4) & $-$0.000017(3) &  $-$0.00001555(9)\\
\hline \\*[-2ex]
\multicolumn{2}{c}{Set Quantities} \\ 
\hline \\*[-2ex]
Epoch of frequency determination (MJD)\dotfill & 53400.0 & 53200.0 & 53200.0 & 53400.0 \\
Epoch of position determination (MJD)\dotfill & 53400.0 & 53200.0 & 53200.0 & 53400.0 \\
Epoch of DM determination (MJD)\dotfill &  53400.0 & 53200.0 & 53200.0 & 53400.0 \\
\hline \\*[-2ex]
\multicolumn{2}{c}{Derived Quantities} \\
\hline \\*[-2ex]
Orbital eccentricity, $e = \sqrt{\epsilon_1^{2}+\epsilon_2^{2}}$\dotfill & - & 1.0(0.4)$\times 10^{-5}$ & 1.9(4)$\times 10^{-5}$ & 1.991(13)$\times 10^{-5}$\\
Omega, $\omega$ = arctan ($\epsilon_1/\epsilon_2)$~($\degr$)\dotfill & - & 114(28) & 152(12) & 218.6(4)\\
Characteristic age, (Gyr)\dotfill & 10.0 & 3.3 & 3.5 & 4.7\\
Surface magnetic field strength, ($10^{8}$G) \dotfill & 1.4 & 13.5 & 22.4 & 4.4\\
Distance$^{b}$, (kpc)\dotfill &  $1.3(2)$ & $1.8(3)$ & $4.0(4)$ & $1.23(13)$\\
Spin period (ms)\dotfill & 3.49663 & 19.3763 & 32.8222 & 7.64587\\
Spin period derivative\dotfill & 5.55$\times 10^{-21}$ & 9.23$\times 10^{-20}$ & 1.47$\times 10^{-19}$  & 2.57$\times 10^{-20}$\\
Mass function, (\msun)\dotfill & $-$ & 0.000591270(4) & 0.00818737(4) & 0.0052494438(8) \\
Minimum companion mass, (\msun)\dotfill & $-$ & 0.11 & 0.28 & 0.24 \\
Total proper motion, $\mu_\mathrm{T}$ (mas\,yr$^{-1}$)\dotfill & 15(25) & 24(2) & 17(37) & 9.2(2)\\
Transverse velocity, V$_\mathrm{T}$ (km\,s$^{-1}$)\dotfill & $<140^{c}$ & 200(50) & $<400^{c}$ & 54(7)\\
\hline \\*[-2ex]
\multicolumn{2}{c}{Assumptions} \\
\hline \\*[-2ex]
Clock correction procedure \dotfill & \multicolumn{2}{c}{TT(TAI)}\\
Solar system ephemeris model \dotfill & \multicolumn{2}{c}{DE405}\\
Model version number \dotfill & \multicolumn{2}{c}{5.00}\\
\hline \\*[-2ex]
\end{tabular}
\\
Note: Figures in parentheses are the nominal 1$\sigma$ \textsc{tempo2}
uncertainties in the least-significant digits quoted.\\ $^{a}$Before
combining the data sets from the different observatories, the errors
on the TOAs were scaled with a constant factor to have each individual
data set return a reduced $\chi^2\approx1$. $^{b}$The DM distances are
estimated from the \cite{cl02} model.  $^{c}$\,Velocity limits from
Eq.\,\ref{eq:pdot}.\\
\end{table*}

We present new timing solutions for the pulsars in
Table\,\ref{tab:solution}, based on 7.5 to 10.5 years of radio pulsar
timing as described in Sect.\,\ref{s:timing}.  The vast majority of
our timing data were taken using a centre frequency around 1380\,MHz,
and therefore originally only the timing variation across the
observing bandwidth was used to measure the dispersion measure (DM) of
the four pulsars. As the combined data sets of WSRT and Nan\c cay or
Jodrell Bank together cover more than 200\,MHz of bandwidth, the DM
was already determined with good accuracy using only that data.
However, to allow for refining the DM value even more, and to study
the pulse shape and flux densities at other frequencies, we also
observed the pulsars at 350\,MHz and 2.3\,GHz. This resulted in a few
detections, see Fig.~\ref{fig:profiles}.  

For PSRs\,J1745$-$0952, J1810$-$2005 and J1918$-$0642, all in very
low-eccentricity binaries, we have used the ELL1 timing model
\citep{lcw+01} to avoid the high degree of covariance between the
epoch and longitude of periastron.

We have measured significant proper motions for PSRs\,J1918$-$0642:
$\mu_\mathrm{T}$\,=\,9.2(2)\,mas\,yr$^{-1}$ and J1745$-$0952:
$\mu_\mathrm{T}$\,=\,24(2)\,mas\,yr$^{-1}$. By combining those proper
motions with the DM-derived distances from the \citealt{cl02} model,
we have calculated transverse velocities for those systems. For the
other two pulsars, we use the marginally significant values from the
timing solution to calculate a limit on their transverse velocities,
see Sect.\,\ref{s:velocities}.

\subsection{Proper motion and velocity}\label{s:velocities}

We use our proper motion measurements to derive transverse velocities,
or limits on those, for all four pulsars (see
Table\,\ref{tab:solution}).  We have a highly significant measurement
of the proper motion of PSR\,J1918$-$0642, and a significant detection
in right ascension for PSR\,J1745$-$0952.  The proper motion values as
presented for the other two pulsars are not significant.
We note that both PSR\,J1721$-$2457 and J1810$-$2005 have
ecliptic latitudes very close to the ecliptic plane ($\beta=-1\fdg8$
and $3\fdg3$ respectively), and therefore their proper motions may
be better and more independently fit in ecliptic parameters. However,
as the coordinate systems are nearly parallel because both
  pulsars are near $\alpha$\,=\,18$^\mathrm{h}$, fits in this
coordinate system did not result in an improvement of the limits on
their proper motions, or the significance of their positions.  As both
of these pulsars lie quite close to the ecliptic, it may be that they
will show effects due to the solar wind \citep{yhc+07a}.  However at
the present timing precision we see no influence on the observed TOAs.

Due to changes in the projected line of sight, transverse motions of
pulsars can affect observed periodicities in the system \citep{shk70}.
We present the contributions to the spin period derivative of each
pulsar in Table\,\ref{tab:pdot}. For completeness, the contributions
to the observed $\dot P$ from accelerations in the Galactic potential
are also presented, although those are not important effects for these
pulsars.
To calculate the magnitude of the Shklovskii term we have used the
total proper motion values as presented in Table\,\ref{tab:solution}.
For PSR\,J1918$-$0642, the contribution to its spin period
derivatives, $\dot P_\mathrm{Shk}$, from proper motion is less than
$8\%$ of the observed value. However for PSR\,J1745$-$0952, the effect
can be as large as $50\%$ of the observed value, leading to an
underestimation of the characteristic age and an overestimation of the magnetic field.
Moreover, as the proper motions for the solitary pulsar J1721$-$2457
and PSR\,J1810$-$2005 have not been determined significantly yet, the
Shlovskii term may even completely dominate the observed $\dot P$ for
those pulsars.
To calculate an upper limit for the proper motion, and accordingly a
limit to the transverse velocity, we can use the observed $\dot P$ as
a maximum:
\begin{equation}\label{eq:pdot}
\left(\frac{\dot P}{P}\right)_\mathrm{Shk} < \left(\frac{\dot P}{P}\right)_\mathrm{Obs}.
\end{equation}
This yields $\mu_\mathrm{T}\,<\,23$\,mas\,yr$^{-1}$ and 
V$_\mathrm{T}\,<\,140$\,\kms for PSR\,J1721$-$2457 and
$\mu_\mathrm{T}\,<\,22$\,mas\,yr$^{-1}$ and V$_\mathrm{T}\,<\,400$\,\kms\ for
PSR\,J1810$-$2005.

About 50\% of the known solitary and low-eccentricity binary pulsars
have measured transverse velocities (Tables\,2 and 4 in
\citealt{lor08}). \cite{hllk05} quote mean values of 2D speeds of
77(16)\,\kms\ for solitary MSPs and 89(15)\,\kms\ for binary
MSPs. Compared to those, the (limits on) transverse velocities that we
have derived from the proper motion measurements represent normal
velocities for recycled pulsars.  

\begin{table}
  \begin{minipage}[t]{\columnwidth}
  \centering
  \caption{Contributions to the period derivative. Contributions due
    to accelerations in the Galactic potential,
    $\dot{P}_\mathrm{Gal}$, are calculated using equations in
    \cite{tsb+99}.
    \label{tab:pdot}}
\begin{tabular}{l@{\hspace{0.30cm}}
c@{\hspace{0.15cm}}
c@{\hspace{0.15cm}}
c@{\hspace{0.15cm}}
c@{\hspace{0.15cm}}}
\hline\hline \\*[-2ex]
PSR &  $\dot{P}_\mathrm{obs}$ & $\dot{P}_\mathrm{Shk,max}$ & $\dot{P}_\mathrm{Gal,\perp}$ & $\dot{P}_\mathrm{Gal,\parallel}$ \\
\hline \\*[-2ex]
J1721$-$2457 & 5.55$\times 10^{-21}$ & $2.5\times 10^{-21}$ & 4.0$\times 10^{-23}$ & 4.4$\times 10^{-22}$ \\
J1745$-$0952 & 9.23$\times 10^{-20}$ & $4.9\times 10^{-20}$ & 4.5$\times 10^{-22}$ & 2.8$\times 10^{-21}$ \\
J1810$-$2005 & 1.47$\times 10^{-19}$ & $5.4\times 10^{-20}$ & 1.1$\times 10^{-23}$ & 1.8$\times 10^{-20}$ \\
J1918$-$0642 & 2.57$\times 10^{-20}$ & $1.9\times 10^{-21}$ & 1.4$\times 10^{-22}$ & 3.7$\times 10^{-22}$ \\
\hline 
\end{tabular}
\end{minipage}
\end{table}

\subsection{Profiles}

\begin{figure*}
  \centering
  \includegraphics[width=12cm, angle=270]{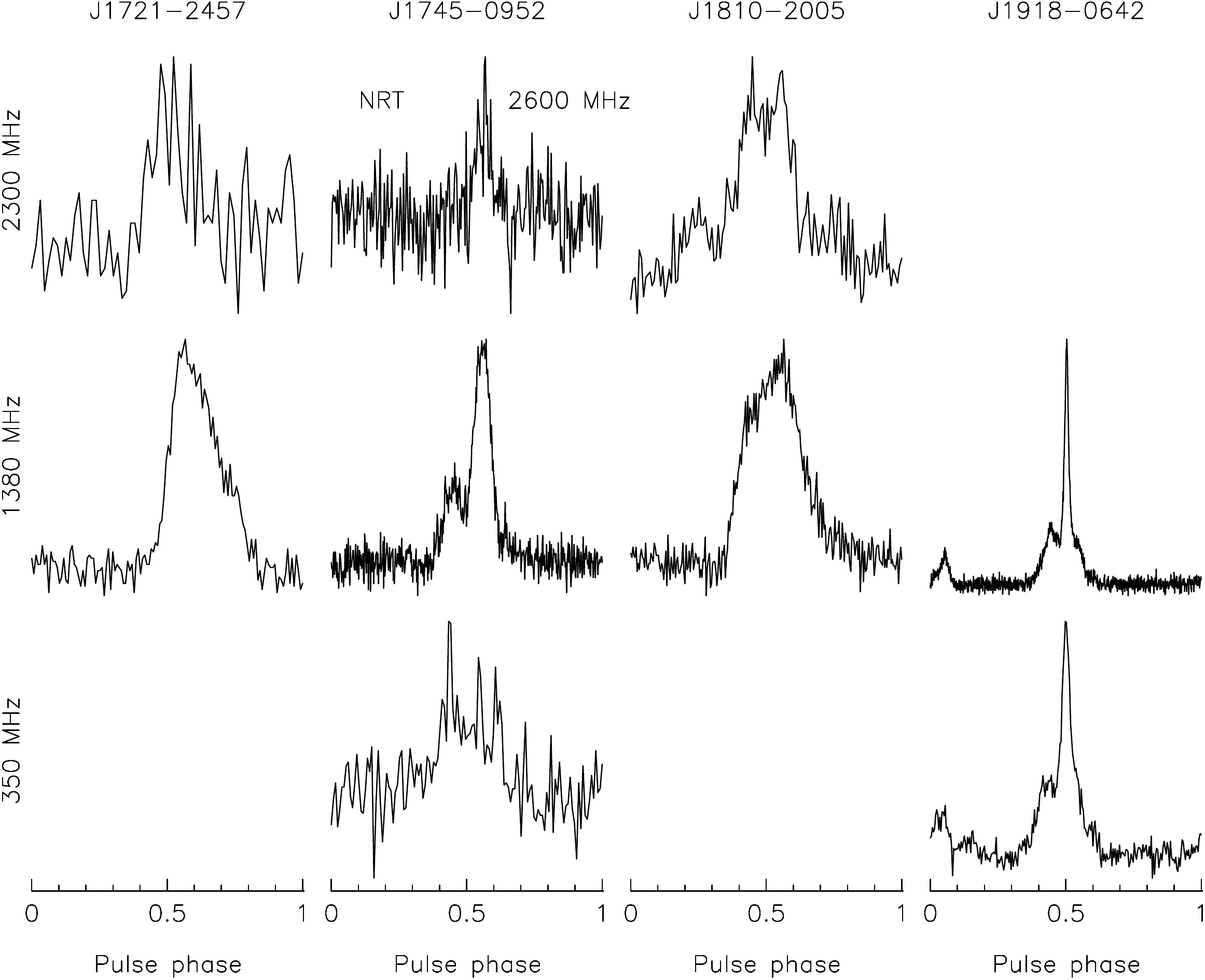} 
  \caption{Profiles for the PSRs at multiple frequencies.
    All profiles were generated from a single 55\,min. observation using
    PuMaII, except the 1380\,MHz profiles of PSRs\,J1721$-$2457,
    J1745$-$0952 and J1810$-$2005 that used 10 to 12 hours of PuMaII
    observations, and the 2600\,MHz profile of PSR\,J1745$-$0952 was
    generated from a 50\,min. Nan\c cay observation.
    These profiles are to show their appearance; for timing
    purposes we used high S/N templates as described in the text.
  \label{fig:profiles}}
\end{figure*}

Apart from our normal timing observing frequency of 1380~MHz, we have
observed all four pulsars at additional frequencies: 350~MHz and
2300~MHz. The results are shown in Fig.~\ref{fig:profiles}. 
It is now standard procedure for WSRT timing observations of MSPs to
use both pulsar machines, and as we have twice the bandwidth
available in PuMaII, we present the profile of the detected pulsars at
350\,MHz and 2.3~GHz from that data, see Fig.~\ref{fig:profiles}.
Single observations of 55~min. were used to generate
the profiles at 350 and 2300~MHz. Where no profile is plotted, the
pulsar was not seen at that frequency. The profiles at 1380~MHz were
generated using one year of regular timing observations and consist of
10 to 12 observations of 25~min, except for PSR\,J1918$-$0642, where
again one observation of 55~min. was used. PSR\,J1745$-$0952 was
detected with NRT at 2600\,GHz in an observation of 50~min.

As reported by \cite{kxl+98}, for most MSPs there is very little
development of pulse profiles with frequency, which can be explained
by MSPs having a very compact magnetosphere.  For all pulsars except
PSR\,J1810$-$2005, we indeed see no changes in the pulse profile with
changing frequency, except for changes in the ratios of the heights of
some components, within the significance of our detections.  Compared
to its profile at 1380~MHz, the 2300~MHz profile of 1810 appears to
have more components in the main peak. Also the profile appears to
show a leading component about 70 degrees before the main peak,
although higher S/N will be needed to confirm this.  Of all the MSP
profiles presented by \cite{kll+99} none show more complex profiles at
higher frequencies than at lower frequencies. Thus a further study of
the profile of PSR\,J1810$-$2005 and its polarisation properties could
reveal more about MSP emission profiles.  The extra features in the
peak of the high-frequency profiles of PSR\,J1810$-$2005 could be an
indication that the profile at 1380\,MHz suffers from scattering
effects. However, the \cite{cl02} model predicts very little
scattering for this pulsar at both 1380~MHz (0.26~ms) and 2300~MHz
(0.03~ms) suggesting that the features in the main peak of the
pulse profile are intrinsic to the pulsar.  It may turn out that, when
better sensitivity can be achieved at higher frequencies,
PSR\,1810$-$2005 will give better timing results compared with the
more commonly used 1380\,MHz as its profile shows more features at
higher frequencies.

\subsubsection{Useful for pulsar timing array?}

PSR\,J1721$-$2457 shows very stable rotational behaviour.
All parameters of the timing solution presented in Table\,\ref{tab:solution} 
are consistent with the original timing solution as published by \cite{eb01b}.
However, for the present S/N ratio, the wide profile of this
pulsar does not allow for the high timing precision 
that is required for PTA pulsars \citep{jhlm05}.

In contrast, despite its relatively long spin period compared to most
pulsars considered for PTA studies, PSR\,J1918$-$0642 may be worthwile
to include. Already at 1380~MHz the profile has a very sharp peak,
allowing for very precise timing over 7.5 years (rms = 2.3\,$\mu$s).
Furthermore, it is possible that better results for pulsar timing
arrays may be deduced by including not only the best-timing pulsars,
but extending the PTA pulsar set with pulsars that have timing
solution with rms $\sim1\,\mu$s \citep{jhlm05,vlml08}.  The pulse
profile as shown in Fig.~\ref{fig:profiles} was generated by the
PuMaII machine in WSRT, promising even better timing precision in the
ongoing timing programme.  Observations with PuMaII use twice the
bandwidth of those with PuMaI and so give an increase in the
sensitivity. Moreover, because PuMaII uses coherent dedispersion, it
yields a sharper profile. The combination of these effects should lead
to an improvement in the timing precision of a factor between 2 and 4
for this pulsar over the next five years which may bring it to the
required level for a PTA pulsar.

\subsection{Flux density and scintillation}\label{s:fluxes}

We calculated flux densities for the four pulsars as described in
Sect.\,\ref{s:wsrt}, based on the profiles as shown in
Fig.\,\ref{fig:profiles}, see Table\,\ref{tab:fluxes}. We note that
apart from those at 21\,cm, the profiles are generated from single
observations only, and therefore the resulting flux densities should
not be regarded as accurate measurements, but are indicative of the
intensities of the pulsars at the observed frequencies.  The spectral
indices for these pulsars are normal except for PSR\,J1810$-$2005,
which appears to have a relatively flat spectrum. We observed this
pulsar a couple of times at 2.3\,GHz, and the resulting flux densities were
varying by a factor 2, indicating that the brightness of this pulsar
may be affected by scintillation.

\begin{figure*}
  \centering
  \includegraphics[width=6cm, angle=270]{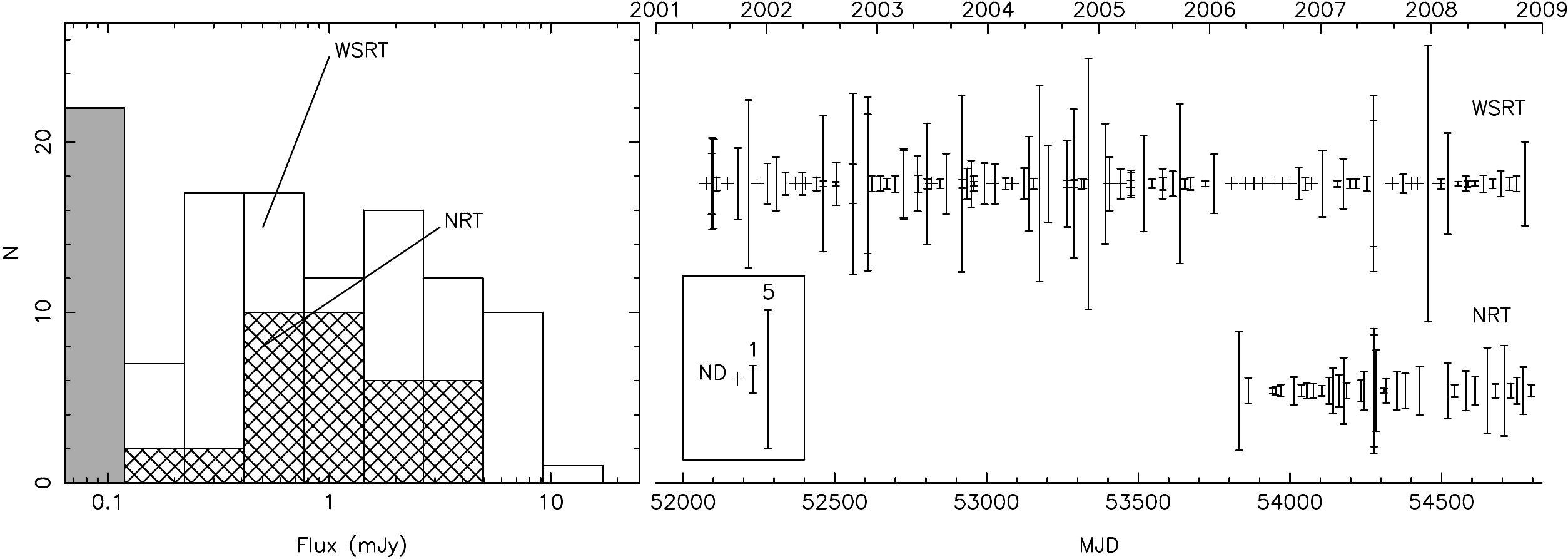}
  \caption{Flux density variations of PSR\,J1918$-$0642 obtained with the WSRT
    and NRT from observations made at 21\,cm. The left panel shows the
    distribution of flux densities from the two observatories, where
    the leftmost column represent all non-detections. In the right
    panel we show the individual flux density values as a function of time,
    where the length of the line in each case is proportional to the
    flux density, as indicated in the legend. Observations where there
    was no clear detection (ND) of the pulsar and thus only an upper
    limit on the flux density can be determined, are plotted with a cross.
    \label{fig:1918flux}}
\end{figure*}

Not long after we began our observations of PSR\,J1918$-$0642 we
noticed that there were a number of occasions when we either did not
detect the pulsar at all or it was extremely bright. A plot of the
flux densities determined from all observations is shown in
Fig.\,\ref{fig:1918flux}.

The left hand panel shows a histogram of flux values from both Nan\c
cay and WSRT observations with the non-detections in the WSRT data
shown in the far left hand bin. The right hand panel shows the
  variation in flux density as a function of time indicating the
  variations are typically on short timescales, although a long period
  of non-detections can be seen in 2006-2007. We note that the pulsar
  is not detected in approximately 20\% of all WSRT observations. The
  pulsar is always detected, although sometimes with very low S/N
  ratio, in the Nan\c cay data. This is not suprising due to the
  higher gain and longer integration times for the Nan\c cay
  observations meaning that the NRT is approximately 30\% more
  sensitive than the WSRT for this pulsar.

If these flux variations are due to scintillation then
the quite frequent non-detections would suggest that the scintillation
bandwidth must be at least as wide as the PuMaI bandwidth of 80\,MHz. 
For the dispersion measure of this pulsar at this observing frequency
such a large scintillation bandwidth seemed too large to be diffractive
\citep{ric01a}. 
Checking the literature, we found four pulsars with
similar dispersion measures which had measured diffractive scintillation
bandwidths in the range 15 -- 70\,MHz (\citealt{jnk98,wmj+05}) at 21\,cm. 
While this is a large range the diffractive scintillation bandwidth would
need to be above the highest measured value to explain the observed
flux variations. 

\begin{figure*}
  \centering
  \includegraphics[width=8.5cm]{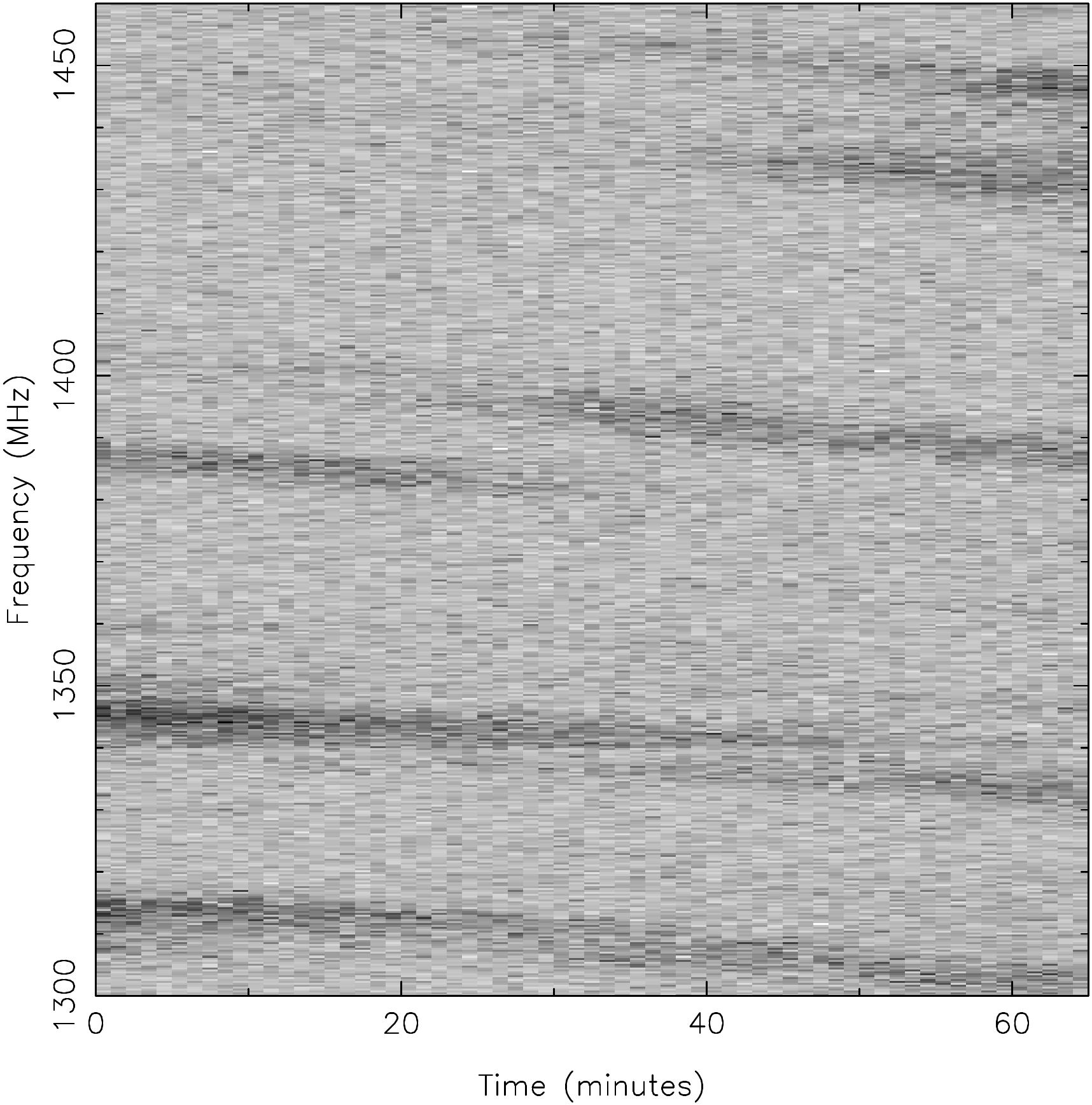}
  \includegraphics[width=8.5cm]{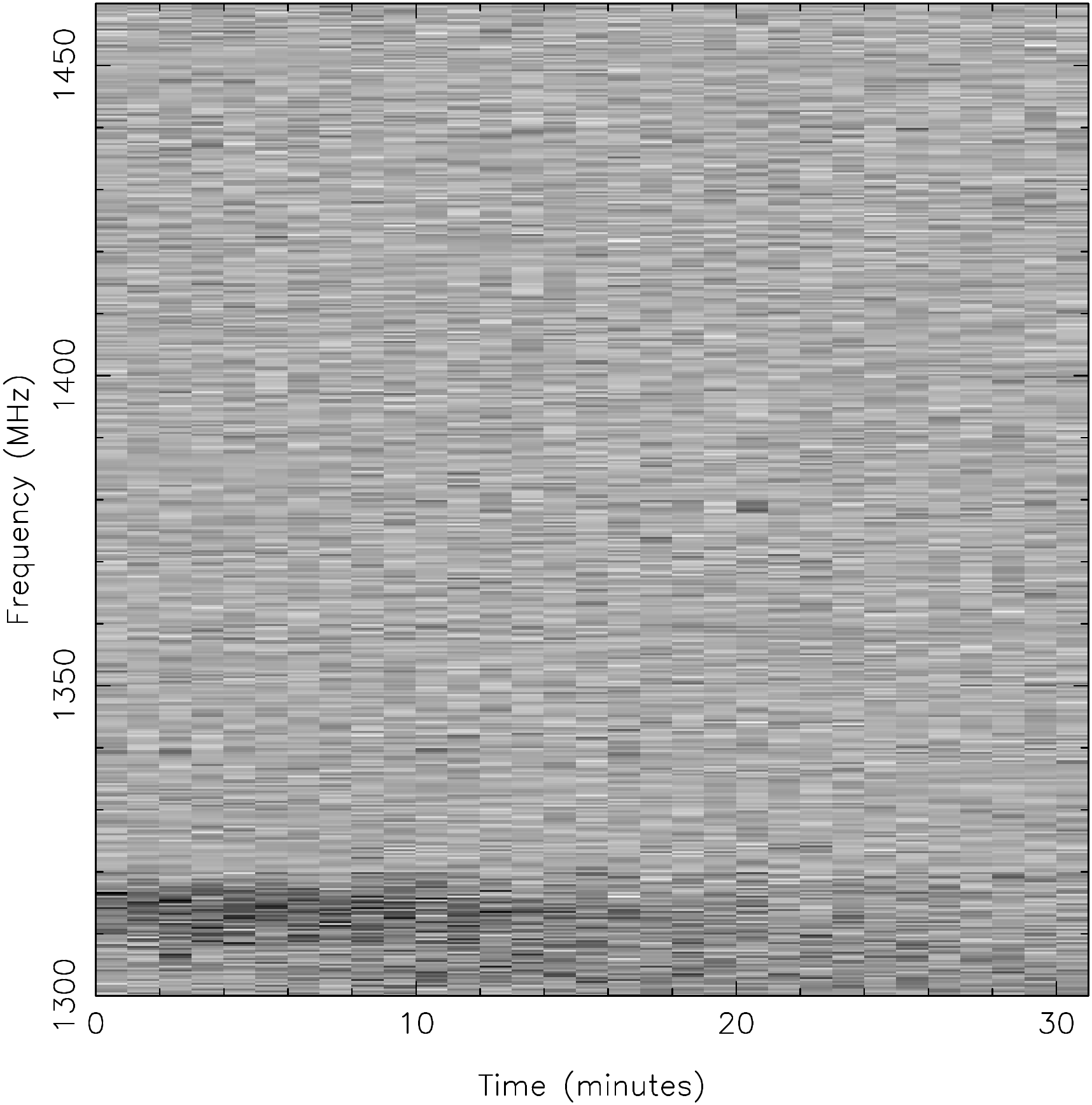} 
  \caption{Dynamic spectra of PSR\,J1918$-$0642 plotted as a greyscale
    of intensity as a function of time and frequency. Darker pixels
    correspond to higher intensity values and the intensity scales
    linearly with the levels of grey. The dynamic spectra are from two
    observations made using PuMaII on 23 February 2008 (left) and 24
    March 2008 (right). 
    \label{fig:1918scintillation}}
\end{figure*}

Using the improved bandwidth of our PuMaII measurements we were able
to better explain the scintillation properties of
PSR\,J1918$-$0642. In Fig.\,\ref{fig:1918scintillation} we show the
dynamic spectra from two observations of PSR\,J1918$-$0642, separated
by just one month.
The left hand side shows a classic case of scintillation 
with a scintillation bandwidth of approximately 5\,MHz. Comparing
this value to the range of bandwidths seen for similar dispersion
measure pulsars, as discussed earlier, it is a factor of three
smaller than the lowest measured scintillation bandwidth.  
In contrast, the right hand plot shows just a single scintle, which is 
perhaps somewhat broader than 5\,MHz, across the full 160\,MHz 
of bandwidth. It may be possible that these variations are purely
statistical in nature, although the lack of any scintiles in the
band of at least 80\,MHz in width, appears to happen about 20\% of
the time. We note that \cite{grl94} 
explain variations in the scintillation properties like this as being 
to the refractive modulation of 
the diffractive interstellar scintillation pattern.
While beyond the scope of this paper, the techniques outlined by 
\cite{grl94} 
could be used to further examine the relationship 
between the velocities in this system and the interstellar medium 
along the line of sight. We note also that very wide bandwidth 
observations of this pulsar provide the best opportunity to do 
high precision timing, as when the pulsar is bright it can be timed 
to high precision.

\begin{table}
  \begin{minipage}[t]{\columnwidth}
  \caption{Flux densities and spectral index.
    \label{tab:fluxes}}
\begin{tabular}{l@{\hspace{0.30cm}}
c@{\hspace{0.15cm}}
c@{\hspace{0.15cm}}
c@{\hspace{0.30cm}}
l@{\hspace{0.15cm}}}
\hline\hline \\*[-2ex]
PSR &  S$_{350}$ (mJy) & S$_{1380}$ (mJy)  & S$_{2273}$ (mJy) & SI \\
\hline \\*[-2ex]
J1721$-$2457 & & 0.58(2) & 0.28(3) & -1.5(2)\\
J1745$-$0952 & 1.8(3) & 0.38(3) & & -1.14(12)\\
J1810$-$2005 & & 1.33(2) & 0.95(4) & -0.67(3)\\
J1918$-$0642 & 5.9(6) & 0.58(2) & & -1.67(6)\\
\hline 
\end{tabular}
\\ *[0.5ex] Note: These numbers correspond to the WSRT profiles in
Fig.~\ref{fig:profiles}. Measurements of flux densities at 350 and
2273\,MHz, and PSR\,J1918$-$0642 at 1380\,MHz are based on one or two
50~min. observations, and therefore should be regarded as
  indicative of the intensities of the pulsars at the observed
  frequencies only. The ranges in flux densities, and consequently,
SI calculations, are now based on the uncertainties in duty cycles.
\end{minipage}
\end{table}

\section{Optical observations}

  \begin{figure*}
    \centering
    \includegraphics[width=8.5cm]{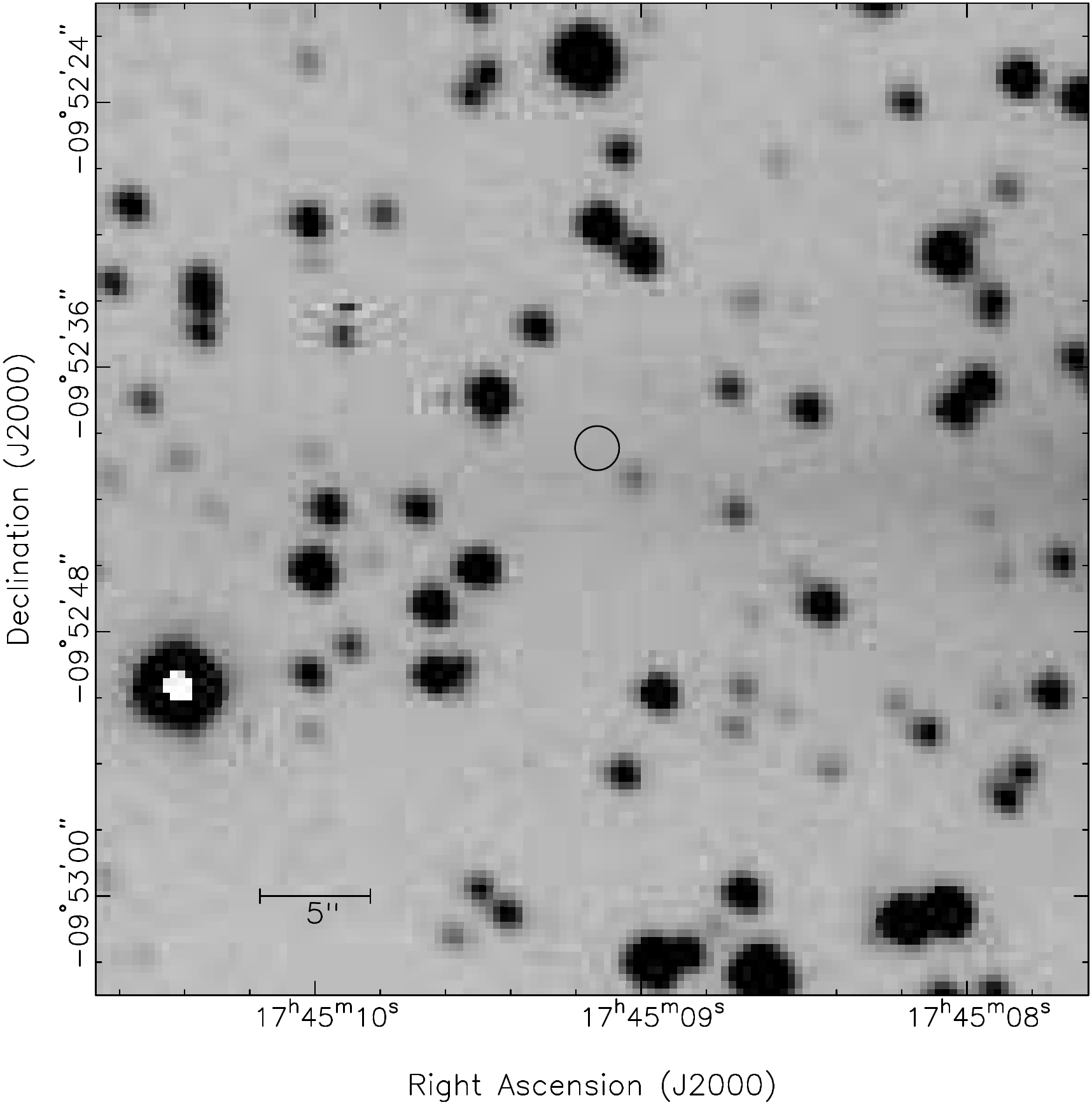}
    \includegraphics[width=8.5cm]{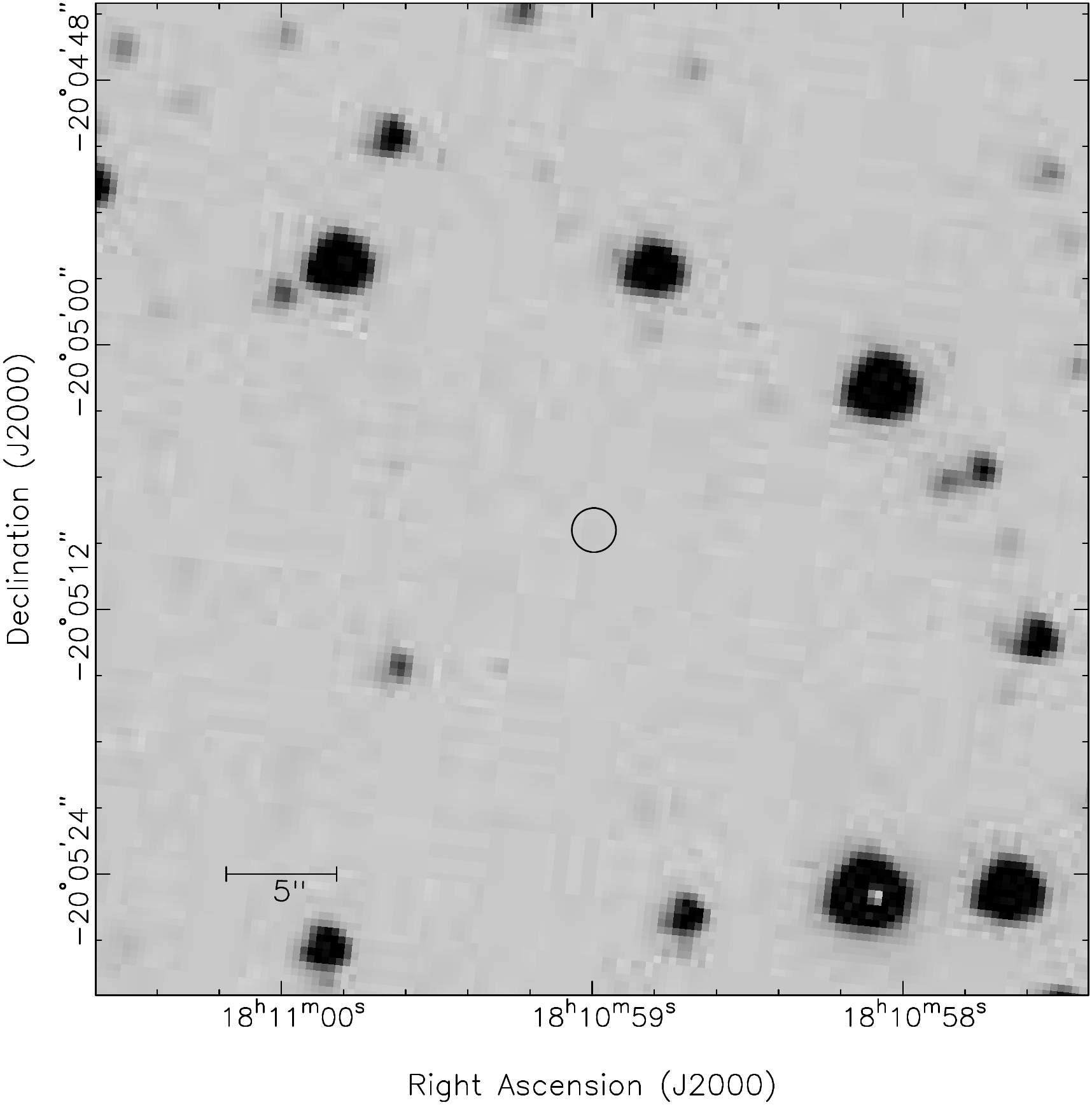}
    \caption{$45\arcsec\times45\arcsec$ subsections of the deep images
      of PSR\,J1745$-$0952 \emph{(left)} and PSR\,J1810$-$2005
      \emph{(right)}. To retain visibility, the pulsar positions are
      denoted by a circles with a $1\arcsec$ radius, even though the
      actual error ellipses are considerably smaller than this.}
    \label{f:finders}
  \end{figure*}

We have analysed archival observations of PSR\,J1745$-$0952 and
PSR\,J1810$-$2005. These observations were obtained with the ESO
Multi-Mode Instrument (EMMI) at the 3.6\,m telescope of the European
Southern Observatory at La Silla, Chile on March 15th, 2004. The
instrument has two 2K$\times$4K detectors 
which were read out with $2\times2$ binning, providing a pixel scale
of $0\farcs33$\,pix$^{-1}$.
The observations consisted of two 10\,min. 
$R$-band exposures of PSR\,J1745$-$0952 and two 5\,min. $R$-band
exposures for PSR\,J1810$-$2005. For both pulsars, these long
exposures were preceded by short (10\,s) exposures, also in the
$R$-band. The seeing during these observations was about $0\farcs86$
to $0\farcs88$. All images were bias-subtracted and flat-fielded using
twilight flats. The two deep images of each pulsar were averaged and
used for further analysis.

The short 10\,s observations were astrometrically calibrated against
the 2nd version of the USNO CCD Astrograph Catalog (UCAC2,
\citealt{zuz+04}). For the PSR\,J1745$-$0952 observations, 34 UCAC2
stars overlapped with the chip on which the pulsar was located, of
which 26 were not saturated and appeared stellar and unblended, giving
rms residuals of $0\farcs056$ in right ascension and $0\farcs091$ in
declination. A secondary astrometric catalog was created from the
stars on the short exposure and used to calibrate the average of the
two 10\,min. exposures. For the transfer, about 300 stars were used,
giving residuals of $0\farcs022$ in right ascension and $0\farcs025$
in declination. Combined with the uncertainty on the timing position
of the pulsar, the quadratic uncertainty of the pulsar position on the
deep optical image is $0\farcs060$ in right ascension and $0\farcs096$
in declination. For PSR\,J1810$-$2005, we used 23 of the 32
overlapping UCAC2 stars, giving rms residuals of $0\farcs049$ in right
ascension and $0\farcs091$ in declination. The transfer of the
secondary catalog to the deeper average of the two 5\,min.
180 secondary standards, giving residuals of $0\farcs028$ in right
ascension and $0\farcs018$ in declination. The final uncertainty on
the pulsar position on the deep image is $0\farcs058$ in right
ascension and $0\farcs220$ in declination. No sources are present in
the error circles of both pulsars, as shown in Fig.\,\ref{f:finders}.

In order to determine upper limits on the brightness of the pulsar
companions, we performed PSF photometry on the averaged deep images
using DAOPHOT\,II \citep{ste87}. The resulting instrumental magnitudes
were calibrated against 47 standard stars in the PG\,1525$-$071 using
the calibrated magnitudes of \citet{ste00}. Since only $R$-band
observations were taken, only the zeropoint was fitted for the
calibration, giving rms residuals of 0.02\,mag. The standard La Silla
$R$-band extinction of 0.07\,mag per airmass was used to correct for
extinction between the standard observations taken at an airmass of
1.08 and the pulsar observations taken at an airmass of 1.34 (for
PSR\,J1745$-$0952) and 1.25 (for PSR\,J1810$-$2005). Based on the
magnitudes and uncertainties of faint stars in the images, we estimate
the $3\sigma$ detection limit at $R>24.49$ for PSR\,J1745$-$0952 and
$R>24.02$ for PSR\,J1810$-$2005.

\subsection{Model predictions}
Though no optical counterparts to PSR\,J1745$-$0952, PSR\,J1810$-$2005
and PSR\,J1918$-$0642 ($R>24$, see \citealt{vbjj05}) were detected, it
is still worthwhile to compare the upper limits on the white dwarf
magnitudes with those predicted by models. To convert the upper limits
on the apparent magnitudes to limits on the absolute magnitude, we
need estimates for the distance and absorption. Estimates for the
distance to each pulsar using the observed dispersion measure and the
\citet{cl02} model for the Galactic distribution of electrons are
given in Table\,\ref{tab:solution}. The $V$-band absorption $A_V$
along the line-of-sight and at the distance of each pulsar was
estimated using the \citet{dcl03} model for Galactic extinction and
converted to $R$-band absorption $A_R$ using the extinction
coefficients of \citet{sfd98} ($A_R=0.819A_V$). We estimate
upperlimits on the absolute $R$-band magnitude of $M_R>11.3$ for
PSR\,J1745$-$0952, $M_R>3.3$ for PSR\,J1810$-$2005 and $M_R>12.7$ for
PSR\,J1918$-$0642.

\begin{figure}
     \resizebox{\hsize}{!}{\includegraphics{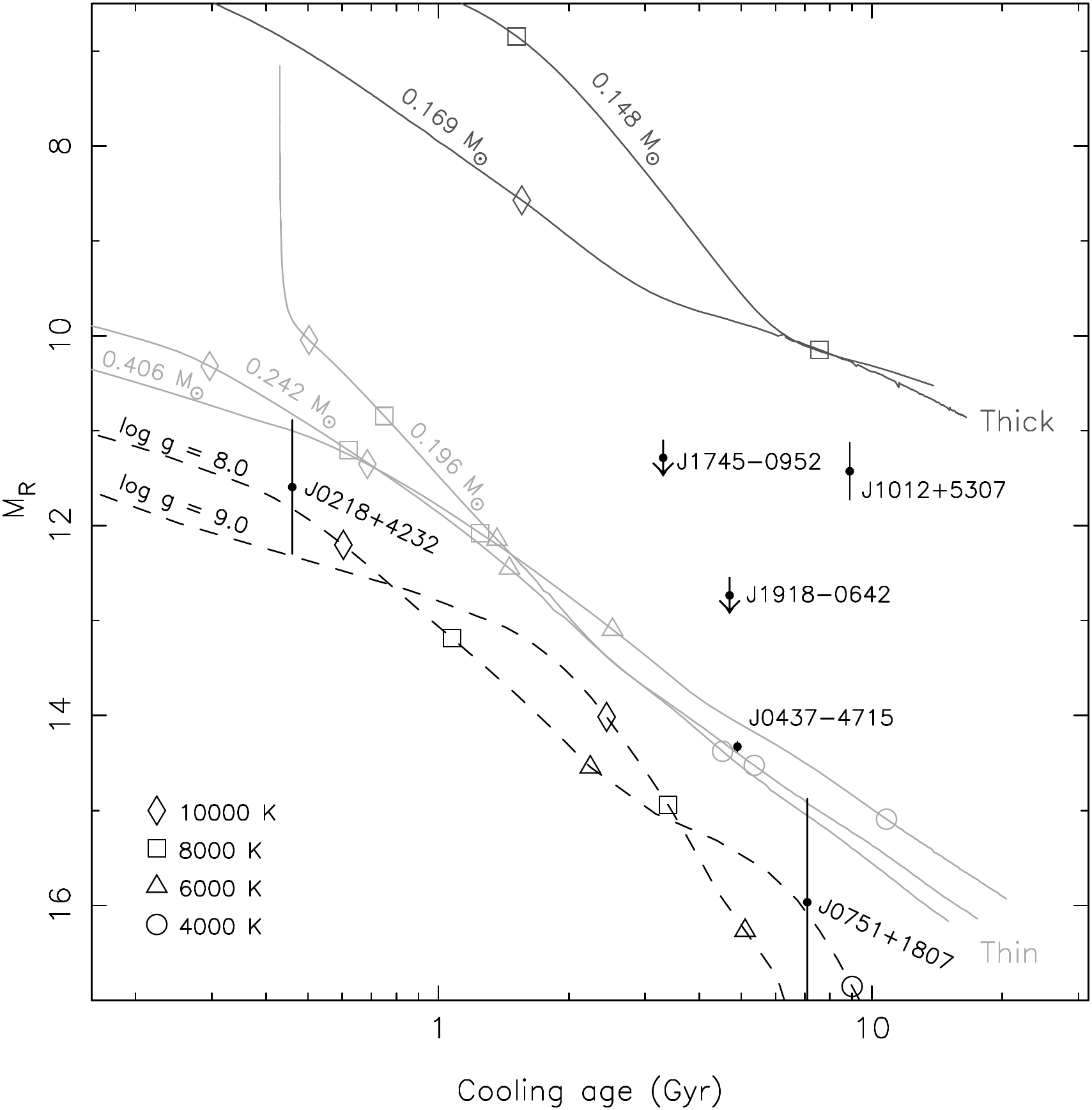}}
     \caption{Absolute $R$-band magnitude and cooling age
       predictions. Shown are helium-core white dwarf cooling tracks
       by \citet{rsab02} and carbon-oxygen white dwarf cooling tracks
       with hydrogen-rich atmospheres by \citet{bwb95} ($\log g=8$
       corresponds to $M_\mathrm{c}\approx0.6$\,M$_\odot$ while $\log
       g=9$ corresponds to
       $M_\mathrm{c}\approx1.2$\,M$_\odot$). Helium-core models with
       masses below approximately 0.18--0.20\,M$_\odot$ have thick
       hydrogen envelopes and continue to residually burn hydrogen,
       keeping the white dwarf hot and slowing the cooling. Heavier
       models have thin hydrogen envelopes and cool much faster,
       creating a dichotomy in the cooling properties of white dwarf
       companions to millisecond pulsars. White dwarfs with
       carbon-oxygen cores cool even faster. Shown with error bars are
       PSR\,J0218+4232 \citep{bvk03}, PSR\,J0437$-$4715 \citep{dbd93},
       PSR\,J0751+1807 \citep{bvk06} and PSR\,J1012+5307
       \citep{llfn95}, where the white dwarf cooling age is assumed to
       be equal to the characteristic age of the pulsar (using braking
       indices of $n=3$). Shown with upper limits on the absolute
       $R$-band magnitude are the values for PSR\,J1745$-$0952 and
       PSR\,J1918$-$0642. For these two pulsars, thick hydrogen
       envelopes are excluded, indicating the white dwarf companions
       in these systems are heavier than about 0.2\,M$_\odot$.}
    \label{f:wdcooling}
  \end{figure}

In Fig.\,\ref{f:wdcooling} we show predictions of absolute magnitude
and cooling ages from the white dwarf cooling tracks by \citet{rsab02}
and \citet{bwb95}.  The helium-core white dwarf cooling tracks show a
dichotomy in the cooling properties of helium-core white dwarfs caused
by differences in the thickness of the hydrogen layer surrounding the
helium core of these white dwarfs (see,
e.g.\,\citealt{sdb00,asb01}). White dwarfs heavier than
0.18--0.20\,M$_\odot$ have thin envelopes and cool faster than lighter
white dwarfs which have thick envelopes. Cooling tracks appropriate
for heavier carbon-oxygen white dwarfs with hydrogen-rich atmospheres are even
fainter.

Under the assumption that the characteristic pulsar ages are equal to
the white dwarf cooling ages, we plot the upper limits on our white
dwarf pulsar companions and detections of others in
Fig.\,\ref{f:wdcooling}. Because of the large distance and absorption
towards PSR\,J1810$-$2005, the limit does not constrain any white
dwarf parameters. For PSR\,J1745$-$0952 and PSR\,J1918$-$0642 the
upper limits exclude white dwarfs with thick hydrogen envelopes. As
such, the mass of the white dwarf in these systems is constrained to
$M_\mathrm{c}\ga0.2$\,M$_\odot$. For PSR\,J1745$-$0952 this mass
limit, combined with the low massfunction of the system, constrains
the inclination of the binary orbit to the low value of $i<34\degr$
(assuming $M_\mathrm{psr}=1.35$\,M$_\odot$). Should this pulsar have a
carbon-oxygen white dwarf, which is not excluded by the optical
observations, the limit on the inclination will be even lower. 

\section{Conclusions} 

Using the WSRT, NRT and Lovell telescopes, we have timed four MSPs for
7.5 to 10.5 years. We have presented updated timing solutions, pulse
profiles at multiple frequencies for each pulsar, and scintillation
parameters for PSR\,J1918$-$0642.  We have measured transverse
velocities for PSRs\,J1745$-$0952 and J1918$-$0642, and set limits on
the velocities of PSRs\,J1721$-$2457 and J1810$-$2005. All velocities
are consistent with previously published distributions for solitary
and binary MSPs.

We have analysed archival optical observations for the binary MSPs and
found no companions to the pulsars.  From the magnitde limits we
deduce for the companions, we can exclude white dwarfs with thick
atmospheres. This indicates that the companions must be heavier than
about 0.2\,\msun.
At this point, the mass restrictions as well as the optical magnitude
limits give no conclusive information to classify PSRs\,J1745$-$0952
or J1810$-$2005 as either LMBPs or IMBPs.

For low-eccentricity binary pulsars, the only post-Keplerian parameter
that is likely to be measurable is the Shapiro delay.  The upper limit
for inclination of PSR\,J1745$-$0952, i$<$34 , suggests that the
effect of Shapiro delay in timing will be very low and therefore we
can not expect to use pulsar timing analyis to disentangle the
individual masses of this system.  However, the expected luminosity of
$M_R>11.3$ indicates that using a dedicated optical observation, the
companion of this pulsar may be detectable, and this will therefore be
the most promising way of deducing the pulsar and companion masses.

\acknowledgements

The Westerbork Synthesis Radio Telescope is operated by ASTRON
(Netherlands Foundation for Research in Astronomy) with support from
the Netherlands Foundation for Scientific Research NWO.  
The Nan\c cay radio Observatory is operated by the Paris Observatory,
associated to the French Centre National de la Recherche Scientifique
(CNRS). The Nan\c cay Observatory also gratefully acknowledges the
financial support of the Region Centre in France.

\end{document}